\documentclass[11pt,reqno]{amsart}
\pdfoutput=1

\newtheorem{theorem}{Theorem}[section]
\newtheorem{proposition}[theorem]{Proposition}
\newtheorem{lemma}[theorem]{Lemma}

\usepackage{helvet,url,cclicenses}
\usepackage{epsf}
\usepackage{graphicx}
\usepackage{ifthen,shortvrb}
\usepackage{array}
\usepackage{latexsym,xspace}
\usepackage{amsmath,amssymb}
\usepackage{stmaryrd}
\usepackage{mathrsfs}
\usepackage{upgreek}
\usepackage{longtable}
\usepackage{algorithm,algorithmic}
\usepackage{color}
\usepackage{tikz}
\usetikzlibrary{shapes}
\usepackage{rotating}
\usepackage{chemarrow}

\DeclareMathAlphabet{\mathpzc}{OT1}{pzc}{m}{it}
\newcommand{\classfont}[1]{\ensuremath{\mathsf{#1}}\xspace}

\newcommand{\NCi}{\classfont{NC^1}}

\newcommand{\ACi}{\classfont{AC^1}}
\newcommand{\Log}{\classfont{L}}

\newcommand{\DLOGCFL}{\classfont{LOGdetCFL}}
\newcommand{\LOGDCFL}{\DLOGCFL}
\newcommand{\p}{\classfont{P}}

\newcommand{\coNP}{\classfont{coNP}}
\newcommand{\PSPACE}{\classfont{PSPACE}}

\newcommand{\ALOGSPACE}[1]{\classfont{ALOGSPACE[\mbox{$ #1$}]}}
\newcommand{\ALOGTIME}{\classfont{ALOGTIME}}
\newcommand{\SPACE}[1]{\classfont{SPACE}(#1)}

\newcommand{\langfont}[1]{\mbox{\textsc{#1}}\xspace}

\newcommand{\ASAGAP}{\langfont{AsAgap}}
\newcommand{\EQRNF}{\langfont{EqRNformula}}

\newcommand{\fe}[1]{\ensuremath{\mbox{\small $#1$}\mbox{-\sc{Mc}}}}

\newcommand{\Klasse}[1]{\ensuremath{\mathcal{#1}}\xspace}

\newcommand{\IL}{\Klasse{IL}}

\newcommand{\Lklasse}[2][]{\ensuremath{\mathrm{#2}#1}\xspace}
\newcommand{\Siv}{\Lklasse[4]{S}}     

\newcommand{\IPC}{\Lklasse{IPC}}
\newcommand{\IPCi}{\ensuremath{\IPC_1}\xspace}

\newcommand{\KC}{\Lklasse{KC}}
\newcommand{\PC}{\Lklasse{PC}}

\newcommand{\imodels}{\models}

\newcommand{\Model}[1]{\ensuremath{\mathcal{#1}}}

\newcommand{\WW}{S}
\newcommand{\inn}{\mathit{in}}
\newcommand{\out}{\mathit{out}}

\newcommand{\Hey}{\mathcal{H}}

\newcommand{\heyrel}{\trianglelefteq}

\newcommand{\muparrow}{\!\uparrow}

\newtheorem{myclaim}{Claim}{\itshape}{\rmfamily}

\newcommand{\ind}{\mathit{RNindex}}
\newcommand{\fib}{\mathit{fib}}
\newcommand{\RNphi}{\mathit{phi}}
\newcommand{\RNpsi}{\mathit{psi}}
\newcommand{\rni}{\mathit{rank}}

\newcommand{\Rphi}{\upvarphi}
\newcommand{\Rpsi}{\uppsi}

\newcommand{\HI}{\mathpzc{h}}

\newcommand{\mlogred}{\leq_{\mathrm{m}}^{\mathrm{log}}}
\newcommand{\VAR}{\operatorname{VAR}}
\newcommand{\PM}{\mathfrak{P}}

\newcommand{\mKlasse}[1]{\mathcal{#1}}

\newlength\problemlength
\newcommand\problemdef[3]{%
\begin{list}{}{\labelwidth\problemlength \labelsep.7em \rightmargin1.5em
\leftmargin\problemlength \advance\leftmargin by3em
\parsep0ex \itemsep.2ex plus.1ex}
\item[{\sl Problem:\hfill}] #1 \item[{\sl Input:  \hfill}] #2
\item[{\sl Output: \hfill}] #3
\end{list}
}
\newcommand\dproblem[3]{%
\vspace{1ex}
\begin{list}{}{\labelwidth\problemlength \labelsep.7em \rightmargin1.5em
\leftmargin\problemlength \advance\leftmargin by3em
\parsep0ex \itemsep.2ex plus.1ex}
\item[{\sl Problem:\hfill}] #1 \item[{\sl Input:  \hfill}] #2
\item[{\sl Question: \hfill}] #3
\end{list}
\vspace{1ex}
}
\newcommand{\apath}{\mathit{apath}}

\newcommand{\qedclaim}{~$\raisebox{0.4mm}{\rule{1.2mm}{1.2mm}}$}

\newcommand{\UPa}{\mathord{\Uparrow}}

\renewcommand{\qed}{\hfill $\Box$}

\newenvironment{mathe}{\vspace{1.2ex}\begin{tabular}{>{$}l<{$}>{$}l<{$}>{$}l<{$}}}{\end{tabular}\vspace{1.2ex}\newline}

\title{The model checking problem for intuitionistic propositional logic with one variable is \ACi-complete}
\author{Martin Mundhenk and Felix Wei\ss}

\begin{document}


\maketitle

\vspace{-2ex}
\begin{small}
\begin{center}
\textsc{Universit\"at Jena, Institut f\"ur Informatik, Jena, Germany} \vspace{0.4ex}\\
\{martin.mundhenk,felix.weiss\}@uni-jena.de
\end{center}
\end{small}

\begin{abstract}
	We  show that the model checking problem 
	for intuitionistic propositional logic with one variable
	is complete for logspace-uniform \ACi.
  As basic tool we use 
  the connection between intuitionistic logic and Heyting algebra,
  and investigate its complexity theoretical aspects.
	For superintuitionistic logics with one variable,
	we obtain \NCi-completeness for the model checking problem.
\end{abstract}


\pagestyle{myheadings}
\thispagestyle{plain}
\markboth{The model checking problem for intuitionistic logic with one variable is \ACi-complete}{Martin Mundhenk and Felix Wei\ss{}}


\section{Introduction}
\label{sec:intro}

Intuitionistic logic (see e.g.~\cite{gabbay81,dal04})
is a part of classical logic
that can be proven using constructive proofs--e.g. by proofs
that do not use \emph{reductio ad absurdum}.
For example, the law of the excluded middle $a \vee \neg a$
and the weak law of the excluded middle $\neg a \vee \neg\neg a$
do not have constructive proofs
and are not valid in intuitionistic logic.
Not surprisingly, constructivism has its costs.
Whereas the validity problem is \coNP-complete for classical propositional logic~\cite{coo71a},
for intuitionistic propositional logic it is \PSPACE-complete~\cite{stat79,svejdar03}.
The computational hardness of intuitionistic logic is already reached 
with the fragment that has only formulas with two variables:
the validity problem for this fragment is already \PSPACE-complete \cite{Rybakov06}.
Recall that every fragment of classical propositional logic with a fixed number of variables
has an \NCi-complete validity problem (follows from~\cite{bus87}).

In this paper, we consider the complexity of intuitionistic
propositional logic \IPC with one variable. 
The model checking problem---i.e. the problem
to determine whether a given formula is satisfied by a given intuitionistic Kripke model---for \IPC is 
\p-complete~\cite{MW-RP10}, even for the fragment with two variables only \cite{MW11}.
More surprisingly, for the fragment with one variable \IPCi
we show the model checking problem to be \ACi-complete.
To our knowledge, this is the first ``natural'' \ACi-complete problem,
whereas formerly known \ACi-complete problems (see e.g.~\cite{bemc95})
have some explicit logarithmic bound in the problem definition.
A basic ingredient for the  \ACi-completeness lies in normal forms for models
and formulas as found by Nishimura \cite{nish60}, that
we reinvestigate under an algorithmic and complexity theoretical point of view. 
In contrast, the formula value problem for classical propositional logic
is \NCi-complete~\cite{bus87} independent of the number of variables.

Classical propositional logic is the extension of \IPC with the axiom $a\vee\neg a$.
Those proper extensions of intuitionistic logic are called superintuitionistic logics.
The superintuitionistic logic \KC (see \cite{DL59}) results from adding $\neg a \vee \neg\neg a$ to \IPC. 
We show that the model checking problem for every superintuitionistic logic with one variable is \NCi-complete
(and easier than that for \IPCi).
In contrast, for the superintuitionistic logic \KC with two variables it is known to be \p-complete
(and as hard as for \IPC with two variables) \cite{MW11}.

As a byproduct, we also obtain results for the validity problem
for intuitionistic and superintuitionistic logics with one variable.

This paper is organized as follows.
In Section~\ref{sec:prelims} we introduce the notations
we use for intuitionistic logic and model checking.
Section~\ref{sec:RNprelims} is devoted to introduce
the old results by Nishimura~\cite{nish60} and to upgrade them
with a complexity analysis.
The following Section~\ref{sec:ac1haerte} presents our lower and upper bound for model checking for \IPCi.
Section~\ref{sec:superint1} deals with the complexity 
of the model checking problem and the validity problem
for superintuitionistic logics with one variable.
The implied completeness for the model checking for intuitionistic logic
and conclusions are drawn in Section~\ref{sec:conclusion}.

 \section{Preliminaries}
\label{sec:prelims}

\subsection*{Complexity (see e.g.\,\cite{vol99})}
    The notion of reducibility we use is the logspace many-one reducibility $\mlogred$,
    except for \NCi-hardness, where we use first-order reducibility.
    \NCi and \ACi are the classes of sets that are decided by
    families of logspace-uniform circuits of polynomial size and logarithmic depth.
    The circuits consist of \mbox{and-}, or-, and not-gates.
    The not-gates have fan-in $1$.
    For \NCi, the and- and or-gates have fan-in $2$ (bounded fan-in),
    whereas for \ACi there is no bound on the fan-in of the gates (unbounded fan-in).
    \ALOGTIME denotes the class of sets decided by alternating Turing machines in logarithmic time,
    and we will use that $\NCi=\ALOGTIME$~(see~\cite{ruz81}).
    \Log denotes the class of sets decidable in logarithmic space.
    We use \ALOGSPACE{f(n)} to denote the class of sets decided by an alternating logspace Turing
    machine that makes $O(f(n))$ alternations, where $n$ is the length of the input.
    We will use that $\ACi=\ALOGSPACE{\log n}$ (see~\cite{coo85}).
    \DLOGCFL is the class of sets that are $\mlogred$-reducible to deterministic context-free languages.
    It is also characterized as the class of sets decidable by 
    deterministic Turing machines in polynomial-time and logarithmic space with additional use of a stack \cite{coo71}.
   The inclusion structure of the classes we use is as follows.
$$
\NCi ~~\subseteq~~ \Log ~~\subseteq~~ \LOGDCFL~~\subseteq~~ \ACi ~~\subseteq~~ \p ~~\subseteq~~ \PSPACE
$$

\subsection*{Intuitionistic Propositional Logic (see e.g.\,\cite{dal04})}
    Let $\VAR$ denote a countable set of \textit{variables}.
    The language \IL of intuitionistic propositional logic 
    is the same as that of propositional logic \PC, i.e. it is the set of all formulas of the form

\begin{mathe}
	\phi & ::= & p \mid \bot \mid (\phi\land\phi) \mid  (\phi\lor\phi) \mid (\phi\rightarrow\phi),
\end{mathe}
    where $p \in \VAR$.
    For $i \geq 0$ the languages  $\IL_i$ are the subsets/fragments
    of \IL for which $\VAR$ consists of $i$ variables.
    In this paper we mainly consider $\IL_1$ (i.e. formulas with one variable).

    As usual, we use the abbreviations  $\neg\phi := \phi\rightarrow\bot$ and $\top := \neg \bot$.
    Because of the semantics of intuitionistic logic, 
    one cannot express $\land$ or $\lor$ using $\rightarrow$ and $\bot$.

An \emph{intuitionistic Kripke model} for intuitionistic logic is a triple $\Model{M} = (U,R,\xi)$,
where $U$ is a nonempty and finite set of \textit{states}, 
$R$ is a preorder on $U$ (i.e. a reflexive and transitive binary relation),
and $\xi: \VAR \to \PM(U)$ is a function%
\footnote{$\PM(U)$ denotes the powerset of $U$.}\,---\,the \textit{valuation function}.
Informally speaking, for any variable it assigns the set of states
in which this variable is satisfied.
The valuation function $\xi$ is monotone in the sense that
     for every $p\in \VAR$, $a,b\in U$: if $a\in \xi(p)$ and $aRb$, then $b\in \xi(p)$.
$(U,R)$ can also be seen as a directed graph. 

    Given an intuitionistic Kripke model $\Model{M} = (U,\leqslant,\xi)$ and a state $s \in U$, the
    \textit{satisfaction relation for intuitionistic logics} $\imodels$ is defined as follows.
    
\begin{mathe}    
  \Model{M},s \not\imodels \bot         			& 								&	\\
  \Model{M},s \imodels p                			& \text{~~iff~~}	& s \in \xi(p),~ p\in\VAR, \\
  \Model{M},s \imodels \phi\land\psi 			& \text{~~iff~~} 	& \Model{M},s \imodels \phi \text{~and~} \Model{M},s \imodels \psi, \\
  \Model{M},s \imodels \phi\lor\psi 				& \text{~~iff~~} 	& \Model{M},s \imodels \phi \text{~or~} \Model{M},s \imodels \psi, \\
  \Model{M},s \imodels \phi\rightarrow\psi & \text{~~iff~~} 	& \forall n\geqslant s : \text{~if~} \mKlasse{M},n \imodels \phi \text{~then~} \mKlasse{M},n \imodels \psi
\end{mathe}
A formula $\phi$ is \textit{satisfied} by an intuitionistic Kripke model $\Model{M}$ in state $s$ if $\Model{M},s \imodels \phi$. 
A \textit{tautology} is a formula that is satisfied by every intuitionistic Kripke model.
Such formulas are also called \textit{valid}.
From the monotonicity of $\xi$ and the definition of $\models$ follows the monotonicity for every formula, i.e. for $\phi \in \IL$, $w,v \in W$ and $w\leqslant v$ if $\Model{M},w \models \phi$, then $\Model{M},v \models \phi$.
\vspace{1ex}

\subsection*{The Model Checking Problem}
    This paper examines the complexity of model checking problems for intuitionistic logics.
  \dproblem{$\fe{\IPC_1}$ }{%
          $\langle \phi,\Model{M},s \rangle$, where $\phi \in \IL_1$, 
              $\Model{M}$ is an intuitionistic Kripke model,
      and $s$ is a state of $\Model{M}$}{$\Model{M},s\imodels\phi$ ?}

We assume that formulas and intuitionistic Kripke models are encoded
in a straightforward way.
This means, a formula is given as a text,
and the graph $(U,R)$ of an intuitionistic Kripke model
is given by its adjacency matrix that takes $|U|^2$ bits.

\section{Properties of \IPCi and its complexity}
\label{sec:RNprelims}

\subsection*{Formulas with one variable}
The set $\IL_1$ of formulas with at most one variable is partitioned into infinitely many equivalence%
\footnote{$\alpha$ is \textit{equivalent} to $\beta$ if every state in every intuitionistic Kripke model satisfies both or none formula. We write $\alpha \equiv \beta$.} classes \cite{nish60}.
This was shown using the formulas that are
inductively defined as follows (see e.g.\cite{gabbay81}).

We use $a$ for the only variable.
\begin{alignat*}{4}
\Rphi_1     & := \neg a                        & \Rpsi_1            			 & := a \\
\Rphi_{n+1} & := \Rphi_{n} \rightarrow \Rpsi_n & \hspace{4ex}  \Rpsi_{n+1} & := \Rphi_{n} \vee \Rpsi_{n} \text{ ~~~~for } n \geq 1
\end{alignat*}

The formulas $\bot,\top,\Rphi_1,\Rpsi_1,\Rphi_2,\Rpsi_2,\ldots$ are called \emph{Rieger-Nishimura formulas}.

\begin{theorem}(\cite{nish60}, cf.\cite[Chap.6,Thm.7]{gabbay81})\label{thm:nish1}
Every formula in $\IL_1$ is equivalent to exactly one of the Rieger-Nishimura formulas.
\end{theorem}

The function $\ind$ maps every formula to the index of its equivalent Rieger-Nishimura formula.
We call this index \textit{Rieger-Nishimura index}.

\begin{mathe}
 \ind(\alpha) & = & \left\{
		\begin{array}{ccl}
			 (i,\RNphi), & \text{~if~} & \alpha \equiv \Rphi_i \\[-3px]
       (i,\RNpsi), & \text{~if~} & \alpha \equiv \Rpsi_i \\[-3px]
       (0,\bot), & \text{~if~} & \alpha \equiv \bot \\[-3px]
       (0,\top), & \text{~if~} & \alpha \equiv \top 
    \end{array}
             \right.
\end{mathe}

\begin{figure}[ht]{
\centering

\begin{minipage}{1\textwidth}
	\centering
	
	\includegraphics[scale=1]{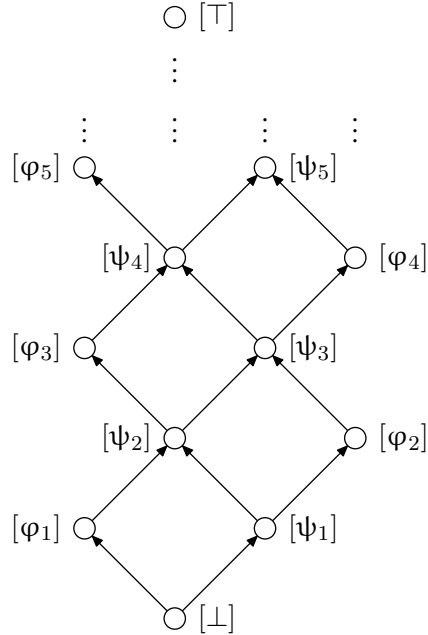}	
\end{minipage}
}

\caption{The Rieger-Nishimura lattice.}
\label{fig:rn-lattice}
\end{figure}

In the following we analyse the complexity of $\ind$.
For $\phi \in \IL_1$ let $[\phi]$ denote the equivalence class that contains $\phi$.
The equivalence classes of $\IL_1$ form a free Heyting algebra over one generator (for algebraic details see~\cite{Johnstone82}).
This algebra is also called the \emph{Rieger-Nishimura lattice} (see Fig.~\ref{fig:rn-lattice}).
It is shown in \cite{nish60} that the lattice operations
can be calculated using a big table look-up (see Appendix~\ref{subsec:RN-lattice-operations}).
For $\alpha,\beta\in\IL_1$, the binary lattice operators $\sqcap$, $\sqcup$ and $\rightarrowtriangle$ are defined as follows.
$[\alpha] \sqcap [\beta] = [\alpha \wedge \beta]$, 
$[\alpha] \sqcup [\beta] = [\alpha \vee \beta]$, and 
$[\alpha] \rightarrowtriangle [\beta] = [\delta]$, 
     where $[\delta]$ is the largest element w.r.t. $\sqsubseteq$%
\footnote{The induced partial order is denoted by $\sqsubseteq$ ($a \sqsubseteq b \Leftrightarrow a \sqcap b = a$).}     
     with $\inf\{[\alpha],[\delta]\} \sqsubseteq [\beta]$.\footnote{$\rightarrowtriangle$ is called the \emph{relative pseudo-complement operation}.}
We use the algebraic properties of \IPCi to give a lower bound on the length of formulas\footnote{$|\alpha|$ 
denotes the length of the formula $\alpha$, 
and it is the number of appearances of variables, connectives, and constants in $\alpha$.}
 in the equivalence classes of $\IL_1$ (Lemma~\ref{lem:formulalength}), and
 to give an upper bound on the complexity of the problem to decide the Rieger-Nishimura index of a formula (Lemma~\ref{lem:RNF_EqTest}).
Let $\rni(\alpha)$ be the first element---the integer---of the $\ind(\alpha)$ pair.

\begin{lemma}\label{lem:formulalength}
For every $\phi \in \IL_1$ it holds that 
$\rni(\phi) \leq c \cdot \log(|\phi|)$, for a constant $c$ independent of $\phi$.
\end{lemma}

\proof
The proof relies on the following technical claim.
Let $\textit{fib}(n)$ denote the $n$-th Fibonacci number%
\footnote{Let $\fib(0)=1$, $\fib(1)=1$, and $\fib(n+2)=\fib(n+1)+\fib(n)$ for $n\geq0$.}.

\begin{myclaim} \label{clm:fib}
Let $\alpha \in \IL_1$. 
Then $|\alpha| \geq \textit{fib}(\rni(\alpha))$.
\end{myclaim}

\emph{Proof of Claim.}
For formulas $\alpha\in[\bot]\cup[\top]$ it holds that $\rni(\alpha)=0$. 
For formulas not in $[\bot]\cup[\top]$,
we prove the claim by induction on the length of $\alpha$.
The only relevant formula of length $1$ is $\alpha=a$.
Since $\rni(a)=1$ and $\fib(1)=1$, the statement holds.

For the induction step let $\alpha \in \IL_1 \setminus ([\bot]\cup[\top])$ with $|\alpha|>1$ and $\alpha = \beta \star \gamma$ with $\star \in \{\rightarrow, \wedge, \vee\}$. 
Then $|\alpha| = |\beta| + |\gamma| +1$, and using the induction hypothesis we obtain $|\alpha| \geq \fib(\rni(\beta)) + \fib(\rni(\gamma)) + 1$. 
We have to distinguish the following cases.
(For the lattice operations see Appendix~\ref{subsec:RN-lattice-operations}.)

\begin{itemize}

\item[\emph{ (i)}] $\gamma \in [\bot]$.
\item[]	Due to the fact that $\alpha \notin [\bot]\cup [\top]$ it follows that $\star \in \{\rightarrow,\vee\}$. 
				If $\star = \vee$, clearly $\beta \in [\alpha]$ and $\rni(\beta)=\rni(\alpha)$. 
				With the induction hypothesis it follows that $|\alpha| \geq \fib(\rni(\beta)) = \fib(\rni(\alpha))$. 
				Otherwise if $\star = \hspace{1ex} \rightarrow$, it follows that $\beta \in [\Rphi_1]\cup [\Rphi_2] \cup [\Rpsi_1]$ and $\alpha \in [\Rphi_2] \cup [\Rphi_1]$. 
				Hence $|\alpha| \geq |\beta| + 2 > 2 > \fib(2) \geq \fib(\rni(\alpha))$.

\item[\emph{ (ii)}] $\beta \in [\bot]$.
\item[] This leads to $\star = \vee$ and can be treated analogously to the case $\gamma \in [\bot]$. 

\item[\emph{ (iii)}] $\beta \in [\top]$ (resp. $\gamma \in [\top]$).
\item[] Remember that $\alpha \notin [\top]$, hence
				independent of the choice of $\star$ it holds that $[\alpha]=[\gamma]$ (resp. $[\alpha]=[\beta]$ and it follows that $|\alpha| > \fib(\rni(\alpha))$.

\item[\emph{ (iv)}] The remaining cases.
\item[] With the induction hypothesis it follows that $|\alpha| \geq \fib(\rni(\beta))+\fib(\rni(\gamma))$.
				With respect to the Rieger-Nishimura lattice, we have to handle two cases.
				\begin{itemize}
					\item[(a)] $\rni(\alpha)\leq \rni(\beta)$ or $\rni(\alpha)\leq \rni(\gamma)$.
					\item[]		 In this case it is not hard to see that $|\alpha| \geq \fib(\rni(\beta))+\fib(\rni(\gamma)) \geq \fib(\rni(\alpha))$.
					\item[(b)] $\rni(\alpha)>\rni(\beta)$ and $\rni(\alpha)>\rni(\gamma)$.
					\item[]		 In this case it holds that one of the ranks of $\beta$ and $\gamma$ needs to be $\geq \rni(\alpha)-2$ and the other $\geq \rni(\alpha)-1$.
										 (See Appendix~\ref{subsec:RN-lattice-operations}, for example $\Rphi_{k-1} \rightarrow \Rpsi_{k-2} \equiv \Rphi_k$ respectively $[\Rphi_{k-1}] \rightarrowtriangle [\Rpsi_{k-2}] = [\Rphi_k]$ for $k \geq 2$.) 
										 Therefore it holds that $|\alpha| \geq \fib(\rni(\alpha)-2) + \fib(\rni(\alpha)-1) = \fib(\rni(\alpha))$. ~~~\qedclaim
				\end{itemize}
\end{itemize}

Claim \ref{clm:fib} shows $|\phi| \geq \textit{fib}(\rni(\phi))$.
Because of the exponential growth of the Fibonacci numbers ($\fib(n) \geq \Phi^n$ where $\Phi$ denotes the golden ratio) 
it follows that $|\phi| \geq c \cdot \log(|\phi|)$ where $c$ is independent of $\phi$.
\qed

In order to analyse the complexity of the Rieger-Nishimura index computation,
we define the following decision problem.

\dproblem{\EQRNF}
				 {$\langle \alpha,(i,x) \rangle$, where $\alpha \in \IL_1$ and $(i,x)$ is a Rieger-Nishimura index}
				 {$\ind(\alpha)=(i,x)$?} 

\begin{lemma}\label{lem:RNF_EqTest}
$\EQRNF$ is in $\DLOGCFL$.
\end{lemma}

\proof
We form Algorithm~\ref{alg:eq_RN_formula} based on the Rieger-Nishimura lattice of the equivalence classes of $\IL_1$. 
The lattice and the lattice operations $\sqcap$, $\sqcup$ and $\rightarrowtriangle$ are described in Appendix~\ref{subsec:RN-lattice-operations}.
We can analogously define the lattice operations $\sqcap$, $\sqcup$ and $\rightarrowtriangle$ for the Rieger-Nishimura indices instead of the equivalence classes%
\footnote{Let $\alpha, \beta, \gamma \in \IL_1$ and $\star \in \{\sqcap,\sqcup,\rightarrowtriangle\}$. 
We set $\ind(\alpha) \star \ind(\beta) = k$ if $[\alpha] \star [\beta] = [\gamma]$ and $k=\ind(\gamma)$.}.

The correctness of Algorithm \ref{alg:eq_RN_formula} is straightforward because the lattice operations for equivalence classes and indices are the same. 
With Lemma \ref{lem:formulalength} it follows that every variable value used in Algorithm \ref{alg:eq_RN_formula} can be stored in logarithmic space.
The algorithm walks recursively through the formula and computes the index of every subformula once, hence running time is polynomial.
All information that are necessary for recursion can be stored on the stack. 
Therefore Algorithm~\ref{alg:eq_RN_formula} can be implemented on a polynomial time logspace machine that uses an additional stack, i.e. a \LOGDCFL-machine.
\qed

\begin{algorithm}[ht]
	\caption{Rieger-Nishimura index check.}
	\label{alg:eq_RN_formula}
  \begin{algorithmic}[1]

		\REQUIRE a formula $\phi \in \IL_i$ and a Rieger-Nishimura index $(i,x)$
		\STATE \textbf{if} RNIndex-calc$(\phi)=(i,x)$ \textbf{then} accept \textbf{else} reject
			\par\vspace{5pt}
			
		\STATE\textbf{function} RNIndex-calc($\psi$) // returns a Rieger-Nishimura index
		\STATE{\textbf{if} $\psi=a$ \textbf{then return} $(1,\RNpsi)$}
		\STATE{\textbf{else if} $\psi=\top$ \textbf{then return} $(0,\top)$}
		\STATE{\textbf{else if} $\psi=\bot$ \textbf{then return} $(0,\bot)$}
		
		\STATE{\textbf{else if} $\psi=\beta \wedge \gamma$ \textbf{then return} RNIndex-calc$(\beta)$ $\sqcap$ RNIndex-calc$(\gamma)$}
		\STATE{\textbf{else if} $\psi=\beta \vee \gamma$ \textbf{then return} RNIndex-calc$(\beta)$ $\sqcup$ RNIndex-calc$(\gamma)$}
		\STATE{\textbf{else if} $\psi=\beta \rightarrow \gamma$ \textbf{then return} RNIndex-calc$(\beta)$ $\rightarrowtriangle$ RNIndex-calc$(\gamma)$}

		\STATE{\textbf{end if}}
	\end{algorithmic}
\end{algorithm}

\subsection*{Canonical models}
\begin{figure}[t]
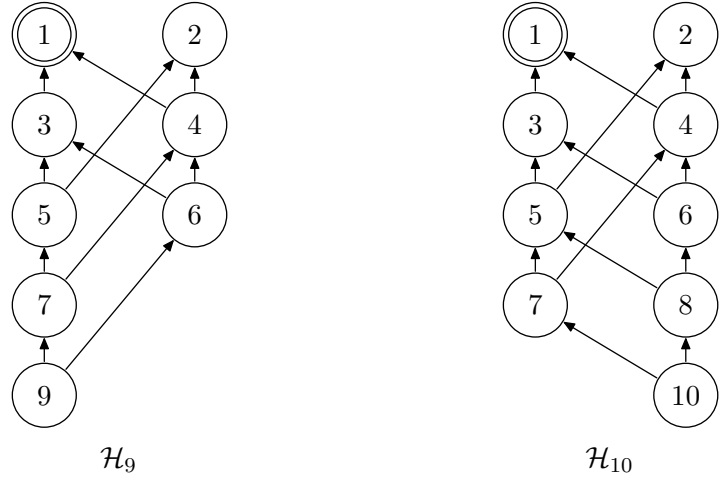
{
\centering

\begin{minipage}{0.4\textwidth}
	\centering
	
	\includegraphics[scale=1]{heyting-modelle.14}

	$\Hey_{9}$
	
\end{minipage}
\begin{minipage}{0.4\textwidth}
	\centering
	
	\includegraphics[scale=1]{heyting-modelle.15}
	
	$\Hey_{10}$
\end{minipage}
}

\caption{The canonical models $\Hey_{9}$ and $\Hey_{10}$
(reflexive and transitive edges are not depicted,
$\xi_n(a)=\{1\}$ is indicated by the double circle for state $1$).}
\label{fig:Heyting models}
\end{figure}

Similar as any formula can be represented by its index, intuitionistic Kripke models can be represented, too.
We give a construction of models---the \textit{canonical models}---
that are also used to distinguish the formula equivalence classes (Theorem \ref{thm:rieger-nishimura}).
Our definition differs a little bit from that in \cite[Chap.6, Defi.5]{gabbay81}.
From Theorems~\ref{thm:rieger-nishimura} and~\ref{lem:hom_heytmodel} it follows that every state $s$ in every intuitionistic Kripke model $\Model{M}$ over one variable has a unique corresponding canonical model $\Hey_n$ in the sense that the state $s$ and the base state%
\footnote{A state is a base state in a model if it has no predecessors (beside itself) w.r.t to the preorder of the model.}
$n$ of $\Hey_n$ satisfy exactly the same formulas.
This was already shown in \cite[Chap.6, Lemma 11]{gabbay81}.
Further define a function $\HI$ that maps $(\Model{M},s)$ to $n$.
For $n\geq 1$, we define the canonical models $\Hey_n=(W_n,\heyrel,\xi_n)$
as follows. 

\begin{mathe}
	W_n & := & \{1,2,\ldots,n-2\}\cup\{n\} \\
	\heyrel &  := & \{ (a,b) \mid  a,b \in W_n, \hspace{2ex} a=b \text{ or } a \geq b+2 \} \\ [+5px]
	\xi_n(a) & := & \begin{cases}
                  				\hspace{1.1ex} \emptyset, & \text{if~} n=2 \\
                  				\{1\}, & \text{otherwise.}
                  \end{cases} 
\end{mathe}
See Figure~\ref{fig:Heyting models} for some examples.

The formulas in $\IL_1$ can be distinguished using the canonical models as follows.

\begin{theorem}(\cite{nish60},cf.\cite[Chap.6, Thm.8]{gabbay81})\label{thm:rieger-nishimura}
For every $n\geq 1$ and every $k \geq 1$ it holds that:
\begin{enumerate}
\item $\Hey_n,n\imodels \Rpsi_k$ ~~iff~~ $n \leq k$
              (i.e. $k\in\{n,n+1,\ldots\}$), ~~and
\item $\Hey_n,n\imodels \Rphi_k$ ~~iff~~ $n < k$ or $n=k+1$
              (i.e. $k\in\{n-1\}\cup\{n+1,n+2,\ldots\}$).
\end{enumerate}
\end{theorem}

For analysing the complexity of the decision problem whether a canonical model is the corresponding model of a state of an arbitrary given intuitionistic Kripke model we define a function $\HI$.
The function $\HI$ maps a given intuitionistic Kripke model $\Model{M}$ and state $w$ of $\Model{M}$ to the index $i$ of the corresponding model $\Hey_i$.
Let $\Model{M}=(W,\leqslant,\zeta)$ be an intuitionistic Kripke model and $w$ a state of $\Model{M}$. 
We define two abbreviations for $w\in W$.

\begin{mathe}
	W_{w\UPa} & := & \{v\in W\mid w\leqslant v\} \\
	W_{w\uparrow} & := & W_{w\UPa} \setminus \{w\}
\end{mathe}
The function $\HI$ is defined as follows.

$$
\HI(\Model{M},w):=\left\{\begin{array}{lll}
    
      1,	 & \mbox{~if} & \mbox{$w\in\zeta(a)$} \\[3px]
      2,	 & \mbox{~if} & \mbox{$w\not\in\zeta(a)$ and $\forall v \in W_{w\muparrow} : v\not\in\zeta(a)$} \\[3px]
      3,	 & \mbox{~if} & \mbox{$w\not\in\zeta(a)$ and $\forall v \in W_{w\muparrow} : \HI(\Model{M},v) \neq 2$ and}  \\
      		 &						& \mbox{$\exists u\in W_{w\muparrow} : \HI(\Model{M},u)=1$} \\[3px]
      n+2, & \mbox{~if} & \mbox{$\forall v\in W_{w\muparrow} : \HI(\Model{M},v) \neq n+1$ and}\\
      		 &						& \mbox{$\exists u_1,u_2 \in W_{w\muparrow} : \HI(\Model{M},u_1)=n$ and $\HI(\Model{M},u_2)=n-1$} \\
    
    \end{array}
   \right.
$$
\noindent
We call $\HI(\Model{M},w)$ the \emph{model index} of $w$ in $\Model{M}$.
The function $\HI$ is well defined because for every state $w$ it holds that 
$\{\HI(\Model{M},v) \mid v \in W_{w\UPa} \} = \{1,2,\dots,\HI(\Model{M},w)-2\} \cup\{\HI(\Model{M},w)\}$.

\begin{theorem}\label{lem:hom_heytmodel}\label{thm:hom_heytmodel}
Let $\Model{M}$ be an intuitionistic Kripke model, $w$ a state of $\Model{M}$, and $k \geq 1$.
Then it holds that
				
\begin{mathe}
	\Model{M},w \models \Rpsi_k & \text{ iff } & k \geq \HI(\Model{M},w) \text{, ~~and} \\
	\Model{M},w \models \Rphi_k & \text{ iff } & k > \HI(\Model{M},w) \text{ ~or~ } k = \HI(\Model{M},w)-1.
\end{mathe}
\end{theorem}

\proof
From Theorem~\ref{thm:rieger-nishimura} follows that (1) is equivalent to the following claim. 

\begin{myclaim}
Let $\Model{M}$ be an intuitionistic Kripke model and $w$ a state of $\Model{M}$.
For every Rieger-Nishimura formula $\alpha$ it holds that $\Model{M},w \models \alpha$ if and only if $\Hey_{\HI(\Model{M},w)},\HI(\Model{M},w) \models \alpha$.
\end{myclaim}

\emph{Proof of Claim.}
We prove this by induction on the rank $\rni(\alpha)$ of $\alpha$.
Let $\Model{M}=(W,\leqslant,\zeta)$ be an intuitionistic Kripke model, $w\in W$ a state, and $\alpha$ a Rieger-Nishimura formula.
The case $\rni(\alpha) \in \{0,1\}$ is clear.
For the induction step we consider a formula $\alpha$ with $\rni(\alpha) > 1$.
We distinguish two cases.
The case $\alpha = \Rpsi_k$ is clear because $\Rpsi_k=\Rphi_{k-1} \vee \Rpsi_{k-1}$ and the claim follows directly from the induction hypothesis.
In the second case we have $\alpha = \Rphi_k$. 

\begin{mathe}
	& \hspace{-6.5ex} \Model{M},w \models \Rphi_k \hspace{2ex} (= \Rphi_{k-1} \rightarrow \Rpsi_{k-1}) & (1) \\
	\Leftrightarrow & \forall v\in W, w \leqslant v : \text{ if } \Model{M},v \models \Rphi_{k-1} \text{ then } \Model{M},v \models \Rpsi_{k-1} & (2) \\
	\Leftrightarrow & \forall v\in W, w \leqslant v : \text{ if } \Hey_{\HI(\Model{M},v)},\HI(\Model{M},v) \models \Rphi_{k-1} & \\[-4px]
	& \hspace{16.5ex} \text{ then } \Hey_{\HI(\Model{M},v)},\HI(\Model{M},v) \models \Rpsi_{k-1} & (3)	\\
	\Leftrightarrow & \forall x\in W_{\HI(\Model{M},w)} : \text{ if } \Hey_x,x \models \Rphi_{k-1} \text{ then } \Hey_x,x \models \Rpsi_{k-1} & (4) \\
	\Leftrightarrow & \forall x\in W_{\HI(\Model{M},w)} : \text{ if } \Hey_{\HI(\Model{M},w)},x \models \Rphi_{k-1} \text{ then } \Hey_{\HI(\Model{M},w)},x \models \Rpsi_{k-1} & (5) \\
	\Leftrightarrow & \Hey_{\HI(\Model{M},w)},\HI(\Model{M},w) \models \Rphi_{k-1} \rightarrow \Rpsi_{k-1} \hspace{2ex} (= \Rphi_k) & (6)
\end{mathe}	
The equivalence between (1) and (2) is clear due to the definition of $\rightarrow$.
From the induction hypothesis follows the equivalence between (2) and (3).
(3) and (4) are equivalent because $\{\HI(\Model{M},v) \mid v \in W,w \leqslant v\} = \{1,2,\dots,\HI(\Model{M},w)-2 \} \cup \{\HI(\Model{M},w)\} = W_{\HI(\Model{M},w)}$.
The definition of the canonical models, i.e. $\Hey_x$ is a submodel of $\Hey_{\HI(\Model{M},w)}$, causes the equivalence between (4) and (5).
The last equivalence between (5) and (6) comes from the definition of $\rightarrow$ and the properties of $\Hey_{\HI(\Model{M},w)}$.~~~\qedclaim
\qed

\section{The complexity of model checking for \IPCi}
\label{sec:ac1haerte}

We first define an \ACi-hard graph problem,
that is similar to the \p-complete alternating graph accessibility problem \cite{chkost81},
but has some additional simplicity properties.
Then we give a construction that transforms 
such a graph into an intuitionistic Kripke model.
This transformation is the basis for the reduction
from the alternating graph accessibility problem to the model checking problem for \IPCi.


\subsection{Alternating graph problems}

The alternating graph accessibility problem is shown to be \p-complete in \cite{chkost81}.
We use the following restricted version of this problem 
that is very similar to Boolean circuits with and- and or-gates (and input-gates).
An \emph{alternating slice graph} \cite{MW-RP10} $G=(V,E)$ is
a directed bipartite acyclic graph with a bipartitioning $V=V_{\exists} \cup V_{\forall}$,
and a further partitioning
$V=V_0 \cup V_1 \cup V_2 \cup \cdots \cup V_{m-1}$ ($m$ \emph{slices}, $V_i\cap V_j=\emptyset$ if $i\not=j$)
where $V_{\exists}=\bigcup_{i<m, i \text{ odd}} V_i$
and $V_{\forall}=\bigcup_{i<m, i \text{ even}} V_i$,
such that $E\subseteq \bigcup_{i=1,2,\ldots,m-1} V_i \times V_{i-1}$ ---
i.e. all edges go from slice $V_i$ to slice $V_{i-1}$ (for $i=1,2,\ldots,m-1$).
All nodes excepted those in the last slice $V_0$ have a positive outdegree.
Nodes in $V_{\exists}$ are called \emph{existential} nodes,
and nodes in $V_{\forall}$ are called \emph{universal} nodes.
Alternating paths from node $x$ to node $y$ are defined as follows by
the property $\apath_G(x,y)$.

\begin{description}
\item[1)] $\apath_G(x,x)$ holds for all $x\in V$
\item[2a)] for $x\in V_{\exists}$: $\apath_G(x,y)$ ~iff~ $\exists z\in V_{\forall}: 
                                                      (x,z)\in E \text{ and } \apath_G(z,y)$
\item[2b)] for $x\in V_{\forall}$: $\apath_G(x,y)$ ~iff~ $\forall z\in V_{\exists}: 
                                          \text{ if } (x,z)\in E \text{ then } \apath_G(z,y)$
\end{description}
The problem \ASAGAP is similar to the alternating graph accessibility problem,
but for the restricted class of alternating slice graphs.
\dproblem{\ASAGAP}{$\langle G,s,t \rangle$, where $G=(V_{\exists}\cup V_{\forall},E)$ is an alternating slice graph
                                                  with slices $V_0,V_1,\ldots,V_{m-1}$,
                                          and $s\in V_{m-1}\cap V_{\exists}$, $t\in V_0\cap V_{\forall}$
}{does $\apath_G(s,t)$ hold?}

Similarly as the alternating graph accessibility problem, \ASAGAP is \p-complete \cite[Lemma 2]{MW-RP10}.
The following technical Lemma is not hard to prove.

\begin{lemma}\label{lemma:ASAGAP-AC1-hard}
For every set $A$ in (logspace-uniform) $\ACi$ exists a function $f$
that maps instances $x$ of $A$ to instances $f(x)=\langle G_x,s_x,t_x \rangle$ of $\ASAGAP$
and satisfies the following properties.
\begin{enumerate}
\item $f$ is computable in logspace.
\item $G_x$ is an alternating slice graph of logarithmic depth;
      i.e.~if $G_x$ has $n$ nodes, then it has $m\leq \log n$ slices.
\item For all instances $x$ of $A$ holds: $x\in A$ if and only if $f(x)\in \ASAGAP$.
\end{enumerate}
\end{lemma}

Essentially, the function $f$ constructs the \ACi circuit $C_{|x|}$ with input $x$,
and transforms it to an alternating slice graph $G_x$.
The goal node $t_x$ represents exactly the bits of $x$ that are $1$.
The start node $s_x$ corresponds to the output gate of $C_{|x|}$,
and $\apath_{G_x}(s_x,t_x)$ expresses that $C_{|x|}$ accepts input $x$.

If we consider $\ASAGAP_{\log}$ as the subset of \ASAGAP
where the slice graphs have logarithmic depth,
this lemma would express that  $\ASAGAP_{\log}$ is $\ACi$-hard under logspace reductions.


\subsection{Alternating slice graphs and intuitionistic Kripke models}
\label{sec:slicegraphs}

Our hardness results rely on a transformation of instances $\langle G,s,t\rangle$ of \ASAGAP to intuitionistic Kripke models $\Model{M}_G:=(U,R,\xi)$.
Let $\langle G,s,t\rangle$ be an instance of \ASAGAP
for the slice graph $G=(V_{\exists}\cup V_{\forall}, E_G)$
with the $m$ slices
$V_{\exists} = V_{m-1}\cup V_{m-3}\cup \cdots \cup V_{1}$
and $V_{\forall} = V_{m-2}\cup V_{m-4}\cup \cdots \cup V_{0}$.

For every $i=0,1,2,\ldots,m-1$, we construct two sets of new states

\begin{mathe}
	W_i^{\inn} & := & \{v^{\inn} \mid v\in V_i\} \text{, ~~and}\\
	W_i^{\out} & := & \{v^{\out} \mid v\in V_i\}
\end{mathe}
and let 

\begin{mathe}
	W & := & \bigcup\limits_{i=0}^{m-1} (W_i^{\inn} \cup W_i^{\out}).
\end{mathe}
Every edge $(u,v)$ from $E_G$ 
is transformed to an edge $(u^{\out},v^{\inn})$
from an $\out$-node to an $\inn$-node,
and every $\inn$-node has an edge to its corresponding $\out$-copy.
This yields the set of edges

\begin{mathe}
	E & := & \big\{(u^{\out},v^{\inn})\mid (u,v)\in E_G\big\} \cup 
                  \big\{(v^{\inn},v^{\out}) \mid v\in V_{\exists}\cup V_{\forall}\big\} ~~.
\end{mathe}
Let $G'=(W,E)$ be the graph obtained in this way from $G$.
If we consider those nodes $v^x\in W$ as $\exists$-nodes (resp. $\forall$-nodes)
that come from nodes $v\in V_{\exists}$ (resp. $v\in V_{\forall}$),
then $\apath_G(u,v)$ if and only if $\apath_{G'}(u^{out},v^{in})$.

\begin{figure}[t]
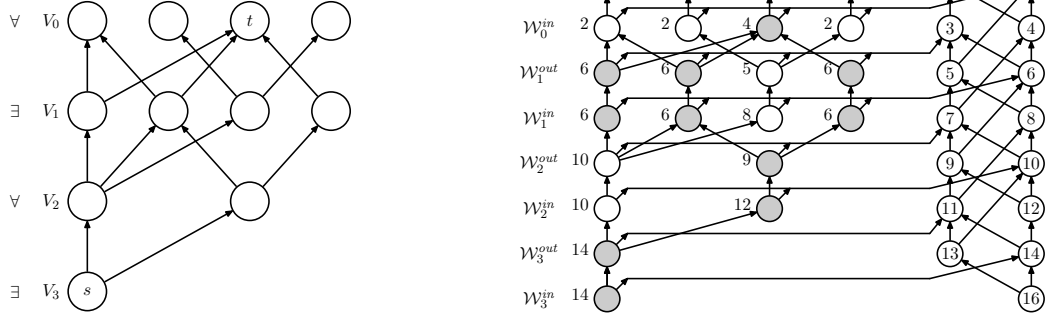

\hfill
\includegraphics[scale=0.6]{agap-example.10}
\hfill\hfill
\includegraphics[scale=0.6]{agap-example.13}
\hfill\mbox{}

\caption{An alternating slice graph $G$ (left) and the resulting intuitionistic Kripke model $\Model{M}_G$ (right);
both the states in $\xi(a)$ are drawn doubly; 
pseudotransitive and reflexive edges in $\Model{M}_G$ are not depicted.
The value at state $x$ denotes its model index $\HI(\Model{M}_G,x)$.
For states in $\Hey_{16}$, their names and their model indices coincide.
States $v^{\inn}$ and $v^{\out}$ for which $\apath_G(v,t)$ holds in $G$ 
are coloured grey.
}
\label{fig:agap-example13}
\end{figure}

Next, we add the nodes of the 
canonical model $\Hey_{4m}=(\{1,2,\ldots,4m-2\}\cup\{4m\},\linebreak \heyrel,\xi_{4m})$ to $G'$ as follows.
Add the nodes $1$ and $2$ to $W_0^{\out}$,
the nodes $3$ and $4$ to $W_0^{\inn}$, the nodes $5$ and $6$ to $W_1^{\out}$ etc.
Formally, for $i=0,1,2,\ldots,m-2$, let 

\begin{mathe}
	\WW_i^{\out} & := & W_i^{\out} \cup \{4 i+1, 4 i+2\}, \\
	\WW_i^{\inn} & := & W_i^{\inn} \cup \{4 i+3, 4 i+4\} \text{, ~~and} \\
	\WW_{m-1}^{\inn} & := & W_{m-1}^{\inn} \cup \{4m\}.
\end{mathe}
The set of states for our model is now 

\begin{mathe}
	U & := & \bigcup\limits_{i=0}^{m-1} (\WW_i^{\out} \cup \WW_i^{\inn}) ~~.
\end{mathe}
Note that $(U,E)$ is still a slice graph 
with slices $\WW_{m-1}^{\inn}, \WW_{m-1}^{\out}, \WW_{m-2}^{\inn}, \ldots$ .
We yet have no edges that connect to nodes from the canonical model.
First we add only those edges between these nodes that do not disturb the ``slice graph'' property,
namely 

\begin{mathe}
	H & := & \{(i,i-2) \mid i\in \{3,4,\ldots,4m-2\}\cup\{4m\}\} \hspace{2ex} \cup \\[-2px]
		&		 & \{(i,i-3) \mid i\in\{4,6,\ldots,4m-2,4m\}\}.
\end{mathe}
Note that $H$ consists of the edges from $\Hey_{4m}$ that give the canonical model its typical structure, i.e. $\heyrel$ is the transitive closure of $H$.
Second we add edges from every node in $W^x_i$ to a node in the neighboured slice
$\WW_{i-1}^{\overline{x}}$ from $\Hey_{4m}$
depending on whether $x=\inn$ or $x=\out$%
\footnote{$\overline{x}=\inn$ if $x=\out$ and vice versa.}.

\begin{mathe}
	T_{\inn} & := & \{ (u,4i+2) \mid u\in W^{\inn}_i, i=0,1,2,\ldots,m-1 \} \\
	T_{\out} & := & \{ (u,4i-1) \mid u \in W^{\out}_i, i=1,2,\ldots,m-1 \}
\end{mathe}
Notice that $(U,E\cup H\cup T_{\inn} \cup T_{\out})$ is still a slice graph
with the slices mentioned above.
It is depicted in Figure~\ref{fig:agap-example13}).
An intuitionistic Kripke model must be transitive and reflexive.
The reduction function that transforms
alternating slice graphs to intuitionistic Kripke models must be computable in logarithmic space.
Within this space bound we cannot compute the transitive closure of a graph.
Therefore, we make the graph transitive with brute force.
We add all edges that jump over at least one slice---we call these edges \emph{pseudotransitive}.\vspace{1.5ex}

\begin{mathe}
	P & := & \bigcup\limits_{i=m-1}^1 \bigg[ \Big(\WW^{\inn}_{i}\times \bigcup\limits_{j=i-1}^0 \WW^{\inn}_{j}\cup\WW^{\out}_{j}\Big) \hspace{2ex} \cup \\
		&		 & \hspace{7.8ex} \Big(\WW^{\out}_{i}\times\big(\WW^{\out}_{i-1}\cup \bigcup\limits_{j=i-2}^0\WW^{\inn}_{j}\cup\WW^{\out}_{j}\big) \Big)
\bigg]
\end{mathe}
Finally, we need to add all reflexive edges.

\begin{mathe}
	T & := & \{(u,u) \mid u\in U\}
\end{mathe}
Notice that the subgraph induced by the states of the canonical model $\Hey_{4m}$
that consists of the edges in $H$ plus the pseudotransitive and the reflexive edges,
is exactly $\Hey_{4m}$.

Eventually, the relation $R$ for our model is 

\begin{mathe}
	R & := & E \cup H \cup T_{\inn} \cup T_{\out} \cup P\cup T,
\end{mathe}
and the valuation function for our model is 

\begin{mathe}
	\xi(a) & := & \{t^{\out},1\},
\end{mathe}
where $t^{\out}$ is the copy of the goal node $t$ in $W_0^{\out}$,
and $\{1\}=\xi_{4m}(a)$ is the node from $\Hey_{4m}$.
This yields the intuitionistic Kripke model $\Model{M}_G = (U, R, \xi)$.
An example of an \ASAGAP instance $\langle G,s,t \rangle$ and the corresponding intuitionistic Kripke model $\Model{M}_G$
constructed from it can be seen in Figure~\ref{fig:agap-example13}.

The states from the canonical model were added to the slice graph
in order to obtain control over the model indices of the other states (w.r.t. the model $\Model{M}_G$).
Our controlling tool is the function $\HI$ which is defined in the previous section.
It maps every state of an intuitionistic Kripke model to its model index.
This is described by Proposition~\ref{claim:h-Werte-der-slices} and Proposition~\ref{claim:knoteneigenschaften}.

\begin{proposition}\label{claim:h-Werte-der-slices}
For every $i=0,1,2,\ldots,m-1$ and every $v\in V_i$ holds
$$
 \HI(\Model{M}_G,v^{\out})\in\{4i+1, 4i+2\} \text{~~~and~~~} \HI(\Model{M}_G,v^{\inn})\in\{4i+2, 4i+4\} ~~.
$$
\end{proposition}

\proof
We prove this by induction on the slices.
For the base case we consider $v\in W_0^{\out}$,
where we have $\HI(\Model{M}_G,v)=1$ if $v=t^{\out}$, and $\HI(\Model{M}_G,v)=2$ if $v\not=t^{\out}$,
and therefore $\HI(\Model{M}_G,v)\in\{1,2\}$.

For the induction step, we consider the remaining slices.

For $v^{\inn}\in W_i^{\inn}$,
we have $(v^{\inn}, 4i+2)\in R$ and $(v^{\out},w)\in R$ for some $w\in W^{\out}_{i}$.
By the induction hypothesis it follows that 
$\HI(\Model{M}_G,u)\leq 4i+2$ for all $u\in U_{v^{\inn}\uparrow}$.
By the definition of $\HI$ it follows that $\HI(\Model{M}_G,v^{\inn}) \in \{4i+2, 4i+4\}$.

For $v^{\out}\in W_i^{\out}$, we have $(v^{\out}, 4i-1)\in R$.
By the induction hypothesis we know that for all $(v^{\out},w)\in R$ with $w\in W^{\inn}_{i-1}$
holds $\HI(\Model{M}_G,w)\in\{4i-2,4i\}$,
and $\HI(\Model{M}_G,u)\leq 4i$ for all $u\in U_{v^{\out}\uparrow}$.
Now,
if for some $w\in W^{\inn}_{i-1}\cap U_{v^{\out}\uparrow}$ holds $\HI(\Model{M}_G,w)=4i$,
then $v^{\out}$ has successors with model indices $4i$ and $4i-1$
and by the definition of $\HI$ it follows that $\HI(\Model{M}_G,v^{\out})=4i+2$.
Otherwise, for all $w\in W^{\inn}_{i-1}\cap U_{v^{\out}\uparrow}$ holds $\HI(\Model{M}_G,w)=4i-2$,
and $v^{\out}$ has no successor with model index $4i$ but successors with model indices $4i-2$ and $4i-1$.
By the definition of $\HI$ it now follows that $\HI(\Model{M}_G,v^{\out})=4i+1$.
\qed

\begin{proposition}\label{claim:knoteneigenschaften}
For every $i=0,1,2,\ldots,m-1$ and every $v\in V_i$ holds:
\begin{enumerate}
\item
    if $i$ is even ($\forall$ slice): \\
    \mbox{}~~~$\apath_G(v,t)$ if and only if $\HI(\Model{M}_G,v^{\out})=4i+1$, ~~and \\
    \mbox{}~~~$\apath_G(v,t)$ if and only if  $\HI(\Model{M}_G,v^{\inn})=4i+4$, \\
\item
    if $i$ is odd ($\exists$ slice):\\
    \mbox{}~~~$\apath_G(v,t)$ if and only if  $\HI(\Model{M}_G,v^{\out})=4i+2$, ~~and \\
    \mbox{}~~~$\apath_G(v,t)$ if and only if  $\HI(\Model{M}_G,v^{\inn})=4i+2$.
\end{enumerate}
\end{proposition}

\proof
We prove this proposition by induction on $i$.
The initial step for $v^{\out}\in W_0^{\out}$ follows directly from the definition of $\Model{M}_G$.

Now for the induction step.
Consider $v\in V_i$ for even $i$ ($\forall$ slice).

\begin{mathe}
	& \hspace{-6.5ex} \apath_G(v,t) & (1) \\
\Leftrightarrow & \forall w\in V_{i-1}, (v,w) \in E_G : \apath_G(w,t) & (2) \\	
\Leftrightarrow & \forall w^{\inn}\in W^{\inn}_{i-1}, (v^{\out}, w^{\inn})\in R : \HI(\Model{M}_G,w^{\inn})=4i-2 & (3) \\
\Leftrightarrow & \{\HI(\Model{M}_G,u) \mid u\in U_{v^{out}\uparrow}\} = \{1,2,\ldots,4i-1\} & (4) \\
\Leftrightarrow & \HI(\Model{M}_G,v^{\out})=4i+1 & (5)
\end{mathe}
(1) and (2) are equivalent by the definition of $\apath_G$.
The equivalence of (2) and (3) comes from the construction of $\Model{M}_G$ and the induction hypothesis.
To show the equivalence of (3) and (4) we prove both the directions separately.
First we show (3) $\Rightarrow$ (4).
Because of (3) there is no $w^{\inn} \in W^{\inn}_{i-1}$ with $\HI(\Model{M}_G,w^{\inn}) > 4i-1$.
If $\{4i-1,4i-2\} \subseteq \{\HI(\Model{M}_G,u) \mid u\in U_{v^{out}\uparrow}\}$, then (4) follows directly.
For $4i-2$ it follows directly from (3).
If $4i-1\not\in \{\HI(\Model{M}_G,u) \mid u\in U_{v^{out}\uparrow}\}$, then $\HI(\Model{M}_G,v^{\out}) \in \{4i\} \cup \{4i-2,4i-3,\dots,1\}$.
Because of (3), it is not possible that $\HI(\Model{M}_G,v^{\out}) < 4i$.
And $\HI(\Model{M}_G,v^{\out})=4i$ is inconsistent with Proposition~\ref{claim:h-Werte-der-slices}.
Hence $\{\HI(\Model{M}_G,u) \mid u\in U_{v^{out}\uparrow}\} = \{1,2,\ldots,4i-1\}$.
For the second direction, (4) $\Rightarrow$ (3), assume that there is some $w^{\inn} \in W^{\inn}_{i-1}$ with $\HI(\Model{M}_G,w^{\inn}) \not= 4i-2$. 
Then from Proposition~\ref{claim:h-Werte-der-slices} it follows that $\HI(\Model{M}_G,w^{\inn})=4i$ but this is inconsistent with (4).
Hence (3) and (4) are equivalent.
(4) equivalent (5) by the construction of $\Model{M}_G$ and the definition of $\HI$.

By Proposition~\ref{claim:h-Werte-der-slices}
we know that $\HI(\Model{M}_G,v^{\out})\in \{4i+1,4i+2\}$.
Remind that $v^{\inn}$ has $v^{\out}$ and $4i+2$ as its direct successors and $\HI(\Model{M}_G,v^{\out})=4i+1$.
Therefore, $\apath_G(v,t)$ if and only if  $\{4i+1,4i+2\}\subseteq \{\HI(\Model{M}_G,u) \mid u\in U_{v^{\inn}\uparrow}\} \subseteq\{1,2,\ldots,4i+2\}$,
where the latter is equivalent $\HI(\Model{M}_G,v^{\inn})=4i+4$.

Finally, we consider $v\in V_i$ for odd $i$ ($\exists$ slice).

\begin{mathe}
	& \hspace{-6.5ex} \apath_G(v,t) & (1) \\
\Leftrightarrow & \exists w\in V_{i-1}, (v,w)\in E_G : \apath_G(w,t) & (2) \\
\Leftrightarrow & \exists w^{\inn}\in W^{\inn}_{i-1}, (v^{\out}, w^{\inn})\in R : \HI(\Model{M}_G,w^{\inn})=4i & (3) \\
\Leftrightarrow & \{4i-1,4i\}\subseteq \{\HI(\Model{M}_G,u) \mid u\in U_{v^{\out}\uparrow}\}\subseteq\{1,2,\ldots,4i\} & (4) \\
\Leftrightarrow & \HI(\Model{M}_G,v^{\out})=4i+2 & (5)
\end{mathe}
(1) and (2) are equivalent by the definition of $\apath_G$.
The equivalence of (2) and (3) comes from the construction of $\Model{M}_G$ and the induction hypothesis.
As in the case above ($i$ is even) the equivalence of (3) and (4) follows from the construction of $\Model{M}_G$ and Proposition~\ref{claim:h-Werte-der-slices}.
(4) equivalent (5) by the construction of $\Model{M}_G$ and the definition of $\HI$.

By Proposition~\ref{claim:h-Werte-der-slices}
we know that $\HI(\Model{M}_G,v^{\out})\in \{4i+1,4i+2\}$.
Remind that $v^{\inn}$ has $v^{\out}$ and $4i+2$ as its direct successors.
Therefore, $\apath_G(v,t)$ 
       if and only if  $\{4i+2\}\subseteq \{\HI(\Model{M}_G,u) \mid u\in U_{v^{\inn}\uparrow}\}\subseteq\{1,2,\ldots,4i\} \cup\{4i+2\}$,
where the latter is equivalent $\HI(\Model{M}_G,v^{\inn})=4i+2$.
\qed

Let $g$ denote the function that maps
instances $x=\langle G,s,t \rangle$ of $\ASAGAP$ to intuitionistic Kripke models $g(x)=\Model{M}_G$ as described above.
The following properties of $g$ are easy to verify.

\begin{lemma}\label{transformation-lemma}
\ \
\begin{enumerate}
\item $g$ is logspace computable.
\item If $x=\langle G,s,t \rangle$ for an alternating slice graph $G$ with $n$ nodes and $m<n$ slices,
      then $g(x)$ is an intuitionistic Kripke model with $\leq 4n$ states and depth $2m$.
\end{enumerate}
\end{lemma}

We will use $g$ as part of the reduction functions for our hardness results.


\subsection{Lower and upper bounds}
\label{sec:AC1 complete}
\label{sec:upper bounds}

Our first result states that the calculation of the
model index of an intuitionistic Kripke model is \p-complete.
It is already \p-complete to decide the last bit of this model index.

\begin{theorem}\label{thm:Hm-Phard}
The following problems are \p-complete.
\begin{enumerate}
\item Given an intuitionistic Kripke model $\Model{M}$ and a state $w$,
decide whether $\HI(\Model{M},w)$ is even.
\item Given an intuitionistic Kripke model $\Model{M}$, a state $w$, and an integer $i$,
decide whether $\HI(\Model{M},w)=i$.
\end{enumerate}
\end{theorem}

\proof
In order to show the \p-hardness of the problems,
we give a reduction from the \p-hard problem \ASAGAP.
From an instance $\langle G,s,t \rangle$ of \ASAGAP
where $G$ is an alternating slice graph with $m$ slices,
construct $\Model{M}=g(\langle G,s,t \rangle)$.
Then $\HI(\Model{M},s^{\out})\in\{4m+1, 4m+2\}$ (Proposition~\ref{claim:h-Werte-der-slices}),
and $\apath_G(s,t)$ if and only if $\HI(\Model{M},s^{out})=4m+2$ (Proposition~\ref{claim:knoteneigenschaften}).
Therefore, $\langle G,s,t \rangle \in \ASAGAP$ if and only if $\HI(\Model{M},s^{out})$ is even
respectively $\HI(\Model{M},s^{out})=4m+2$.

For every intuitionistic Kripke model $\Model{M}=(U,\leqslant,\xi)$ it holds that $\HI(\Model{M},w)\leq|U|+1$. 
To decide for a given intuitionistic Kripke model $\Model{M}$, a state w of $\Model{M}$, and an integer $n$ the problem ``Does $\HI(\Model{M},w)=n$ hold?'' is in \ALOGSPACE{n}.
The function $\HI$ can be implemented according to its definition straightforwardly as a logarithmically space bounded alternating algorithm.
It requires an alternation depth of at most $n$ due to the construction of $\HI$.
Using that $\p=\ALOGSPACE{\mathit{poly}}$ \cite{chkost81} then it follows that both problems are in \p.
\qed

In the construction of the above proof,
the decision whether $\HI(\Model{M},s^{out})=4m+2$ is the same
as to decide whether $\Model{M},s^{out} \imodels \Rpsi_{4m+2}$,
for the Rieger-Nishimura formula $\Rpsi_{4m+2}$ (Theorems~\ref{thm:rieger-nishimura} and \ref{lem:hom_heytmodel}).
Unfortunately, the length of $\Rpsi_{4m+2}$ is exponential in $m$ (Lemma~\ref{lem:formulalength}),
and therefore the mapping from $\langle G,s,t \rangle$ (with $m$ slices) 
to the model checking instance $\langle \Rpsi_{4m+2}, g(\langle G,s,t \rangle), s^{out} \rangle$
cannot in general be performed in logarithmic space.
But if the depth $m$ of the slice graph is logarithmic,
the respective formula $\Rpsi_{4m+2}$ has polynomial size only and the reduction works in logarithmic space.

\begin{theorem}\label{thm:IPC1-is-ACi-hard}
The model checking problem for $\IPCi$ is \ACi-hard.
\end{theorem}

\proof
Let $B$ be in $\ACi$.
By Lemma~\ref{lemma:ASAGAP-AC1-hard} there exists a logspace computable function $f_B$
such that for all instances $x$ of $B$, $x\in B$ if and only if $f_B(x)\in \ASAGAP$,
where $f_B(x)=\langle G_x, s_x, t_x \rangle$
for an alternating slice graph $G_x$ with $n_x$ nodes and $m_x\leq \log n_x$ slices.
The following function $r$ reduces $B$ to the model checking problem for \IPCi.

\begin{mathe}
	r(x) & = & \langle \Rpsi_{4m_x+2}, g(f_B(x)), s_x^{\out} \rangle
\end{mathe}
\indent
\emph{$r$ can be computed in logspace.}
Since $f_B$ is logspace computable,
it follows that  $g(f_B(x))$ and $s_x^{\out}$ can be computed in logspace.
The Rieger-Nishimura formula $\Rpsi_{4m_x+2}$ can also be computed in logspace,
because $m_x$ is logarithmic in $|x|$ and therefore 
 $\Rpsi_{4m_x+2}$ has length polynomial in $|x|$.

\emph{$B$ logspace reduces to the model checking problem for \IPCi via the reduction function $r$.}
By Propostion~\ref{claim:knoteneigenschaften} we have that $\langle G_x,s_x,t_x \rangle\in \ASAGAP$ if and only if \linebreak $\HI(g(\langle G_x,s_x,t_x \rangle),s_x^{\out})=4m_x+2$.
By the properties of the Rieger-Nishimura formulas (Theorem~\ref{thm:rieger-nishimura})
this is equivalent to $g(\langle G_x,s_x,t_x \rangle), s_x^{\out} \imodels \Rpsi_{4m_x+2}$.
This shows the correctness of the reduction.
\qed

In the following theorem we show an upper bound for the \IPCi model checking problem.

\begin{theorem}\label{thm:AC1-algorithm} 
The model checking problem for \IPCi is in \ACi.
\end{theorem}

\proof
First we show that Algorithm \ref{alg:AC1_MC} decides the model checking problem and then we analyse its complexity.

We show that Algorithm \ref{alg:AC1_MC} accepts the input $\langle \varphi, \Model{M}, s \rangle$ if and only if $\Model{M},s \imodels \varphi$.
Informally speaking Algorithm \ref{alg:AC1_MC} accepts the input if and only if $\ind(\varphi)$ and the model index $\HI(\Model{M},s)$ of $s$ in $\Model{M}$ match according to Theorem~\ref{lem:hom_heytmodel}.

Instead of computing the equivalent Rieger-Nishimura formula, Algorithm \ref{alg:AC1_MC} only calculates its Rieger-Nishimura index. 
This is done in Lines 1 and 2. 
The trivial cases are handled in Lines 3 and 4.
From Theorem~\ref{lem:hom_heytmodel} we know for an arbitrary Rieger-Nishimura formula $\alpha_k$ with $\rni(\alpha_k)=k>0$ the following.
Either $\alpha_k\!=\!\Rpsi_k$ and it holds that $\HI(\Model{M},s)\leq k$ if and only if $\Model{M},s\imodels \alpha_k$. 
This is checked in Line 6.
Or $\alpha_k\!=\!\Rphi_k$ and it holds that $\HI(\Model{M},s)=k+1$ or $\HI(\Model{M},s)<k$ if and only if $\Model{M},s\imodels \alpha_k$.
This is checked in Line 9.
If $\HI(\Model{M},s) > \rni(\varphi)+1$, then it holds that $\Model{M},s \not \imodels \varphi$ (Theorems~\ref{thm:rieger-nishimura} and~\ref{lem:hom_heytmodel}).

In the following, we estimate the complexity of Algorithm \ref{alg:AC1_MC}.
It gets $\langle \varphi,\Model{M},s \rangle$ as input.
In Line 1 Algorithm~\ref{alg:AC1_MC} guesses a Rieger-Nishimura index $(r,x)$.
The decision in Line 2 whether $\langle \varphi,(r,x) \rangle \in \EQRNF$ can be done with the resources of \DLOGCFL (Lemma~\ref{lem:RNF_EqTest}).
To decide for a given intuitionistic Kripke model $\Model{M}$, a state w of $\Model{M}$, and an integer $n$ the problem ``Does $\HI(\Model{M},w)=n$ hold?'' is in \ALOGSPACE{n}.
The function $\HI$ can be implemented according to its definition straightforwardly as a logarithmically space bounded alternating algorithm.
It requires an alternation depth of at most $n$ due to the construction of $\HI$.
Hence the decision in Line 6 (resp. Line 9) whether $\HI(\Model{M},s) \in \{1,2, \dots, r\}$ (resp. $\HI(\Model{M},s) \in \{1,2, \dots, r-1\} \cup \{r+1\}$) can be done with $r$ (resp. $r+1$) alternations. 
Since $r$ is at most about $c \cdot \log(|\phi|)$ (Lemma~\ref{lem:formulalength}),
these decisions can be done with at most $c \cdot \log (|\langle \phi, \Model{M}, s\rangle|)$ alternations.
During the complete computation, the algorithm only needs to store a constant number of Rieger-Nishimura indices and model indices.
According to Lemma \ref{lem:formulalength} and the fact that $\HI(\Model{M},w)\leq |\Model{M}|$,
Algorithm \ref{alg:AC1_MC} requires during the alternations logarithmic space.
Since $\LOGDCFL\subseteq\ACi=\ALOGSPACE{\log n}$, we obtain the desired upper bound.
\qed

\begin{algorithm}[t]
	\caption{model checking algorithm for $\IPCi$}
	\label{alg:AC1_MC}
  \begin{algorithmic}[1]

		\REQUIRE a formula $\phi\in \IL_1$, an intuitionistic Kripke model $\Model{M}$ and a state $s$
		
		\STATE {guess nondeterministically a Rieger-Nishimura index $(r,x)$ with $r \leq c \cdot \log(|\phi|)$ } 
		\STATE {\textbf{if} $\langle \phi,(r,x) \rangle \in \EQRNF$ \textbf{then} }
				
		\STATE {\hspace{2ex}\textbf{if} $(r,x)=(0,\bot)$ \textbf{then} reject}
		\STATE {\hspace{2ex}\textbf{else if} $(r,x)=(0,\top)$ \textbf{then} accept}
		
		\STATE{\hspace{2ex}\textbf{else if} $x=\RNpsi$ \textbf{then}}
			\STATE {\hspace{4ex}\textbf{if} $\HI(\Model{M},s) \in \{1,2, \dots, r\}$ \textbf{then} accept}
			\STATE {\hspace{4ex}\textbf{else} reject}

		\STATE{\hspace{2ex}\textbf{else if} $x=\RNphi$ \textbf{then}}
			\STATE {\hspace{4ex}\textbf{if} $\HI(\Model{M},s) \in \{1,2, \dots, r-1\} \cup \{r+1\}$ \textbf{then} accept}
			\STATE {\hspace{4ex}\textbf{else} reject}
		
		\STATE{\hspace{2ex}\textbf{end if}}
		\STATE{\textbf{else} reject}
	\end{algorithmic}
\end{algorithm}


\section{Some notes on superintuitionistic logics with one variable}
\label{sec:superint1}

Superintuitionistic propositional logics are logics that
have more valid formulas than \IPC.
In this sense, classical propositional logic is a superintuitionistic
logic, since it can be obtained as the closure under substitution and modus ponens of 
the tautologies from \IPC plus $a\vee\neg a$ as additional axiom.
A well-studied superintuitionistic logic is \KC \cite{DL59}
that results from adding the weak law of the excluded middle $\neg a \vee\neg\neg a$ to \IPC.
Semantically, the intuitionistic Kripke models for \KC are restricted to those intuitionistic Kripke
models $\Model{M}=(W,\leqslant,\xi)$ where $\leqslant$ is a {directed} preorder.
Whereas $\IL_1$ over preorders has infinitely many equivalence classes of formulas,
$\IL_1$ over directed preorders has only 7 equivalence classes---represented by the 
Rieger-Nishimura formulas $\bot,\top,\Rphi_1,\Rpsi_1,\Rphi_2,\Rpsi_2,\Rphi_3$---that can be distinguished using the
first 3 canonical models~\cite{nish60,mak66}.
This follows from $\neg a \vee \neg\neg a \equiv \Rpsi_3$.
The function $\HI$ can be implemented for such models as an alternating Turing machine
that runs in logarithmic time, if the function value is fixed to a finite range---that 
in this case is $\{1,2,3\}$---independent of the input.
For $\KC_1$, the Rieger-Nishimura index of the formulas
also has a finite range (as mentioned above).
Therefore, it can be calculated by an alternating Turing machine that runs in logarithmic time
similar to the machine presented by Buss~\cite{buss93}
that calculates the value of a Boolean formula.
Instead of the Boolean values $0$ and $1$,
here we have $7$ different Rieger-Nishimura indices.
The rules how the index of a formula can be calculated
from the indices of its subformulas and the connective,
follow directly from the Rieger-Nishimura lattice operations---see Appendix~\ref{subsec:RN-lattice-operations}.
If the indices are bound to a finite range,
this big table yields an even bigger but finite table without index-variables.
For example, the equivalence $\Rphi_n \vee \Rphi_{n+1} \equiv \Rpsi_{n+2}$ for all $n\geq 1$
induces the three equivalences $\Rphi_1 \vee \Rphi_{2} \equiv \Rpsi_{3}$,
$\Rphi_2 \vee \top \equiv \top$, and $\top \vee \top \equiv \top$ for $\KC_1$.
This yields alternating logarithmic-time ($=\NCi$) as upper bound for the validity problem for $\KC_1$.

There are infinitely many superintuitionistic logics (with one variable) that can be obtained by adding
any not valid formula as axiom to \IPCi.
For example, if we add a formula equivalent to $\Rpsi_k$,
then the superintuitionistic logic obtained has finitely many equivalence classes
represented by $\bot,\top,\Rphi_1,\Rpsi_1,\ldots, \Rphi_{k-1}, \Rpsi_{k-1}, \Rphi_k$.
With similar arguments as for $\KC_1$ we can conclude that
the model checking problems of these logics all are in \NCi.
Moreover, the formula value problem for Boolean formulas without variables
is \NCi-hard~\cite{bus87}.
Intuitionistic formulas without variables have the same values,
if they are interpreted as classical Boolean formulas.
This means, the semantics of $\rightarrow$
is the same for Boolean formulas and for intuitionistic formulas without variables.
Therefore, the model checking problem for any superintuitionistic logic without variables is \NCi-hard, too.

The validity problem for superintuitionistic logic has the same complexity,
since in order to decide whether a formula with one variable is valid
it suffices to know its Rieger-Nishimura index.

 \section{Conclusion}
\label{sec:conclusion}


We consider computational problems that appear
with intuitionistic propositional logic without variables and with one variable.
We characterize the complexity of model checking for intuitionistic logic.

\begin{theorem}\label{theo:conclusion}
\ \
\begin{enumerate}
\item\label{part1} The model checking problem for $\IPC_0$ is \NCi-complete.
\item\label{part2} The model checking problem for $\IPC_1$ is \ACi-complete.
\end{enumerate}
\end{theorem}

Part(\ref{part1}) follows from the fact that an intuitionistic formula that contains
constants $\bot$ and $\top$ but no variables
can be evaluated like a Boolean formula,
whose evaluation problem is \NCi-complete~\cite{bus87} independently
of the number of variables.
Part~(\ref{part2}) follows from Theorems~\ref{thm:IPC1-is-ACi-hard} and \ref{thm:AC1-algorithm}.
It shows a difference between \IPCi and its modal companion \Siv with one variable,
for which the model checking problem is \p-complete~\cite{MW-RP10}.

Intuitionistic logic with one variable turns out to be very interesting.
There are infinitely many equivalence classes of formulas,
and according to Lemma~\ref{lem:formulalength}
even the sequence of smallest formulas of these equivalence classes has an exponential growth with respect to the length of the formulas.
Such a fast growing sequence seems to appear rarely in ``natural'' problems,
and it is a key ingredient for the
\ACi-completeness of the model checking problem.
Intuitionistic logic with one variable
is strongly related to free Heyting algebras with one generator.
Since Heyting algebras are generalizations of Boolean algebras,
it would be interesting to investigate whether the difference between \NCi and \ACi
is related to that between Boolean algebras and Heyting algebras.

\begin{theorem}\label{theo:super-int-mc}
The model checking problem for every superintuitionistic logic with one variable is \NCi-complete.
\end{theorem}

This follows from the discussion in Section~\ref{sec:superint1}.

It is interesting to notice that 
the complexity results for \IPC and for $\KC$ with at least two variables are the same for the model checking problem \cite{MW11}.
But for the fragments with one variable, the complexity of \IPCi is higher
than that of $\KC_1$.

The fragments of \IPC with a restricted number of variables and $\rightarrow$ as only connective 
have finitely many equivalence classes of formulas and models~\cite{Urq74,RHJ10}.
The equivalence class of a given formula can be obtained with the resources of $\NCi$, using a technique from Buss~\cite{bus87}.
This might indicate an upper bound lower than \p for the model checking problem.
For the implicational fragment with at most one variable,
$\NCi$-completeness follows from Theorem~\ref{sec:superint1}.
But a general result for an arbitrary number of variables is open.


For the validity problem we obtain the following results.

\begin{theorem}\label{thm:validity_conclusion}
\ \
\begin{enumerate}
	\item\label{cpart1} The validity problem for every superintuitionistic logic with one variable is \NCi-complete.
	\item\label{cpart2} The validity problem for $\IPCi$ is in $\SPACE{\log n \cdot \log\log n}\cap \LOGDCFL$.
\end{enumerate}
\end{theorem}

Part~(\ref{cpart1}) follows from the discussion in Section~\ref{sec:superint1}.
Part~(\ref{cpart2}) is from Svejdar \cite{svejdar_personal09} and Lemma~\ref{lem:RNF_EqTest}.
The exact complexity of the validity problem for \IPCi is open.
It is interesting to notice that
superintuitionistic logics with one variable all have lower complexity than \IPCi,
whereas for superintuitionistic logics with two variables already \KC reaches the same complexity as \IPC (follows from Rybakov~\cite{Rybakov06}).


If we consider other problems related to Kripke models for \IPCi
that are not ``out braked'' by a very fast growing part of the input,
the complexity jumps up to \p-complete\-ness, as shown in Theorem~\ref{thm:Hm-Phard}.
Model checking for \IPCi also gets \p-hard
if the instances $\langle \varphi, \Model{M}, s \rangle$
allow the formula $\varphi$ to be represented as a graph.
This holds even for formulas without variables, 
and therefore it also holds for all superintuitionistic logics.
If formulas are represented as graphs, the 
sequence of smallest representatives of the equivalence classes of \IPCi
does not have exponential growth anymore.
Moreover, the calculation of the Rieger-Nishimura index gets \p-hard.

\begin{theorem}\label{theo:P-complete-formulas}
If the formulas are represented as graphs, the following problems are \p-complete:
\begin{enumerate}
\item\label{ppart1} the model checking problem for \IPCi,
\item\label{ppart1a} the model checking problem for every superintuitionistic logic with one variable,
\item\label{ppart2} the validity problem for \IPCi, and
\item\label{ppart2a} the validity problem for every superintuitionistic logic with one variable.
\end{enumerate}
\end{theorem}

Parts~(\ref{ppart1}) and (\ref{ppart1a}) contrast the different upper bounds 
\NCi and \ACi 
for the standard encodings of formulas (Theorem~\ref{theo:super-int-mc} resp. Theorem~\ref{theo:conclusion}).
Parts~(\ref{ppart2}) and (\ref{ppart2a}) contrast the complexity of
the validity problems for the logics under consideration (Theorem~\ref{thm:validity_conclusion}).


\vspace{4ex}

\textbf{Acknowledgements.}
The authors thank Vitek Svejdar, Heribert Vollmer,
and Thomas Schneider for helpful discussions. 

\vspace{4ex}

\textbf{Remark.}
This work is an extended version of~\cite{MW-STACS11}.
Theorem 15 in~\cite{MW-STACS11} and its proof can be found in~\cite[Theorem 3.6]{MW11}.


\bibliographystyle{abbrv}

\newpage 
\appendix

\section{The Rieger-Nishimura lattice operations}
\label{subsec:RN-lattice-operations}

\noindent
Let $[\varphi]$ denote the equivalence class that contains $\varphi$, for being $\varphi \in \IL_1$.
The equivalence classes of $\IL_1$ form a free Heyting algebra over one generator (see~\cite{gabbay81}).
This algebra is also called the \emph{Rieger-Nishimura lattice} (see Fig.~\ref{fig:rn-lattice}),
and is given by $(\{a\},\sqcap,\sqcup,\rightarrowtriangle,\bot)$ whereas $a$ denotes the only one variable that occurs in the formulas of $\IL_1$. 
The induced partial order is denoted by $\sqsubseteq$ ($a \sqsubseteq b \Leftrightarrow a \sqcap b = a$).
For $\alpha,\beta\in\IL_1$, the binary lattice operators $\sqcap$, $\sqcup$ and $\rightarrowtriangle$ are defined as follows.
$[\alpha] \sqcap [\beta] = [\alpha \wedge \beta]$, 
$[\alpha] \sqcup [\beta] = [\alpha \vee \beta]$, and 
$[\alpha] \rightarrowtriangle [\beta] = [\delta]$, 
     where $[\delta]$ is the largest element w.r.t. $\sqsubseteq$ with $\inf\{[\alpha],[\delta]\} \sqsubseteq [\beta]$.\footnote{$\rightarrowtriangle$ is called the \emph{relative pseudo-complement operation}.}
In~\cite{nish60} the following properties of the operations of the Rieger-Nishimura lattice (see Figure \ref{fig:rn-lattice}) are shown.
We describe these properties as equivalences of Rieger-Nishimura formulas.
This is very similar to \cite[Chap.6,Thm.7]{gabbay81}. 
For example because of $\Rphi_{n+1} \rightarrow \Rpsi_n \equiv \Rphi_{n+2}$ it holds that $[\Rphi_{n+1}] \rightarrowtriangle [\Rpsi_n] = [\Rphi_{n+2}]$.
\newline \vspace{3ex}

\begin{small}
\begin{tabular}{ll}
$\Rphi_n \rightarrow \Rphi_n \equiv \top$ & $\Rphi_n \vee \Rphi_n \equiv \Rphi_n$\\
$\Rphi_n \rightarrow \Rphi_{n+1} \equiv \Rphi_{n+1}$ & $\Rphi_n \vee \Rphi_{n+1} \equiv \Rpsi_{n+2}$\\
$\Rphi_n \rightarrow \Rphi_{n+k} \equiv \top$ for $k>1$ & $\Rphi_n \vee \Rphi_{n+k} \equiv \Rphi_{n+k}$ for $k>1$ \\
$\Rphi_{n+k} \rightarrow \Rphi_n \equiv \Rphi_n$ for $k \geq 1$ \hspace{4ex} & $\Rphi_n \vee \Rpsi_n \equiv \Rpsi_{n+1}$ \\ 
$\Rphi_n \rightarrow \Rpsi_n \equiv \Rphi_{n+1}$ & $\Rphi_n \vee \Rpsi_{n+k} \equiv \Rpsi_{n+k}$ for $k \geq 1$ \\
$\Rphi_n \rightarrow \Rpsi_{n+k} \equiv \top$ for $k\geq1$ & $\Rphi_{n+k} \vee \Rpsi_n \equiv \Rphi_{n+k}$ for $k \geq 1$ \\
$\Rphi_{n+1} \rightarrow \Rpsi_n \equiv \Rphi_{n+2}$ & $\Rpsi_n \vee \Rpsi_m \equiv \Rpsi_{\max \{n,m\}}$ \\
$\Rphi_{n+2} \rightarrow \Rpsi_n \equiv \Rphi_{n+1}$ & $\Rphi_n \vee \bot \equiv \Rphi_n$ \\
$\Rphi_{n+k} \rightarrow \Rpsi_n \equiv \Rpsi_n$ for $k>2$ & $\Rpsi_n \vee \bot \equiv \Rpsi_n$ \\
$\Rpsi_n \rightarrow \Rpsi_n \equiv \top$ & $\Rphi_n \vee \top \equiv \top$ \\
$\Rpsi_n \rightarrow \Rpsi_{n+k} \equiv \top$ for $k\geq1$ & $\Rpsi_n \vee \top \equiv \top$ \\
$\Rpsi_{n+1} \rightarrow \Rpsi_n \equiv \Rphi_{n+1}$ & $\bot \vee \top \equiv \top$ \\
$\Rpsi_{n+k} \rightarrow \Rpsi_n \equiv \Rpsi_{n}$ for $k>1$ & $\bot \vee \bot \equiv \bot$ \\
$\Rpsi_n \rightarrow \Rphi_n \equiv \Rphi_n$ & $\top \vee \top \equiv \top$ \\
$\Rpsi_{n+k} \rightarrow \Rphi_n \equiv \Rphi_n$ for $k \geq 1$ & $\Rphi_n \wedge \Rphi_n \equiv \Rphi_n$ \\
$\Rpsi_n \rightarrow \Rphi_{n+k} \equiv \top$ for $k \geq 1$ & $\Rphi_1 \wedge \Rphi_2 \equiv \bot$ \\
$\Rphi_1 \rightarrow \bot \equiv \Rphi_2$ & $\Rphi_n \wedge \Rphi_{n+1} \equiv \Rpsi_{n-1}$ for $n>1$ \\
$\Rphi_2 \rightarrow \bot \equiv \Rphi_1$ & $\Rphi_n \wedge \Rphi_{n+k} \equiv \Rphi_n$ for $k>1$ \\
$\Rphi_n \rightarrow \bot \equiv \bot$ for $n>2$ & $\Rphi_1 \wedge \Rpsi_1 \equiv \bot$ \\
$\Rphi_n \rightarrow \top \equiv \top$ & $\Rphi_n \wedge \Rpsi_n \equiv \Rpsi_{n-1}$ for $n>1$ \\
$\Rpsi_1 \rightarrow \bot \equiv \Rphi_1$ & $\Rphi_n \wedge \Rpsi_{n+k} \equiv \Rphi_n$ for $k \geq 1$ \\
$\Rpsi_n \rightarrow \bot \equiv \bot$ for $n>1$ & $\Rphi_{n+k} \wedge \Rpsi_n \equiv \Rpsi_n$ for $k \geq 1$ \\
$\Rpsi_n \rightarrow \top \equiv \top$ & $\Rpsi_n \wedge \Rpsi_m \equiv \Rpsi_{\min \{ n,m \}}$ \\
$\top \rightarrow \Rphi_n \equiv \Rphi_n$ & $\Rphi_n \wedge \bot \equiv \bot$ \\
$\bot \rightarrow \Rphi_n \equiv \top$ & $\Rpsi_n \wedge \bot \equiv \bot$ \\
$\top \rightarrow \Rpsi_n \equiv \Rpsi_n$ & $\Rphi_n \wedge \top \equiv \Rphi_n$ \\
$\bot \rightarrow \Rpsi_n \equiv \top$ & $\Rpsi_n \wedge \top \equiv \Rpsi_n$ \\
$\bot \rightarrow \top \equiv \top$ & $\bot \wedge \top \equiv \bot$ \\
$\top \rightarrow \bot \equiv \bot$ & $\bot \wedge \bot \equiv \bot$ \\
$\bot \rightarrow \bot \equiv \top$ & $\top \wedge \top \equiv \top$ \\
$\top \rightarrow \top \equiv \top$ & 
\end{tabular}
\end{small}

\end{document}